# Self-organizing memristive nanowire networks with structural plasticity emulate biological neuronal circuits

Gianluca Milano[1,2], Giacomo Pedretti[3], Matteo Fretto[2], Luca Boarino[2], Fabio Benfenati[4,5], Daniele Ielmini*[3], Ilia Valov*[6,7], Carlo Ricciardi*[1]

[1]Department of Applied Science and Technology, Politecnico di Torino, C.so Duca degli Abruzzi 24, 10129 Torino, Italy.

[2]Advanced Materials Metrology and Life Science Division, INRiM (Istituto Nazionale di Ricerca Metrologica), Strada delle Cacce 91, 10135 Torino, Italy.

[3]Dipartimento di Elettronica, Informazione e Bioingegneria, Politecnico di Milano and IU.NET, Piazza L. da Vinci 32, 20133, Milano, Italy.

[4]Center for Synaptic Neuroscience and Technology, Istituto Italiano di Tecnologia, Largo Rosanna Benzi, 10, 16132 Genova, Italy.

[5]IRCCS Ospedale Policlinico San Martino, Largo Rosanna Benzi, 10, 16132 Genova, Italy.

[6]JARA – Fundamentals for Future Information Technology, 52425 Jülich, Germany.

[7]Peter-Grünberg-Institut (PGI 7), Forschungszentrum Jülich, Wilhelm-Johnen-Straße, 52425 Jülich, Germany.

*e-mails: daniele.ielmini@polimi.it; i.valov@fz-juelich.de; carlo.ricciardi@polito.it;

**Acting as artificial synapses, two-terminal memristive devices are considered fundamental building blocks for the realization of artificial neural networks[1–3]. Organized into large arrays with a top-down approach, memristive devices in conventional crossbar architecture demonstrated the implementation of brain-inspired computing for supervised and unsupervised learning[4]. Alternative way using unconventional systems consisting of many interacting nano-parts have been proposed for the realization of biologically plausible architectures where the emergent behavior arises from a complexity similar to that of biological neural circuits[5–9]. However, these systems were unable to demonstrate bio-realistic implementation of synaptic functionalities with spatio-temporal processing of input signals similarly to our brain. Here we report on emergent synaptic behavior of biologically inspired nanoarchitecture based on self-assembled and highly interconnected nanowire (NW) networks realized with a bottom up approach. The operation principle of this system is based on the mutual electrochemical interaction among memristive NWs and NW junctions composing the network and regulating its connectivity depending on the input stimuli. The functional connectivity of the system was shown to be responsible for heterosynaptic plasticity that was experimentally demonstrated and modelled in a multiterminal configuration, where the formation of a synaptic pathway between two neuron terminals is responsible for a variation in synaptic strength also at non-stimulated terminals. These results highlight the ability of nanowire memristive architectures for building brain-inspired intelligent systems based on complex networks able to physically compute the information arising from multi-terminal inputs.**

Cognitive functions of humans stem from the emergent behavior of biological neural networks composed of ~ $10^{14}$-$10^{15}$ synaptic connections in between neurons, whereby the high connectivity of the system provides robustness, adaptability and fault tolerance[10]. Synaptic connections undergo rapid changes in synaptic strength in response to activity and recent history of the neuron (short-term plasticity) that shape information processing within the network[11]. These short-lived changes in synaptic weights can evolve in long-term changes that rely on alterations of synaptic connections, impacting the architecture and topology of the neural hardware. In addition to input-specific Hebbian changes in active synapses (homosynaptic plasticity), plasticity can be induced also at a larger population of synapses that were not active during the induction of Hebbian plasticity (heterosynaptic plasticity) and contributes to the stability and homeostasis of neural networks[12].

Taking inspiration from the recurrent connectivity of biological systems (Fig. 1a), memristive devices based on highly interconnected NWs were realized by drop-casting Ag-NWs in suspension on a $SiO_2$ insulating substrate. Subsequent patterning of Au electrodes proceeds (Fig. 1b and Extended Data Fig. 1) without need of cleanroom facilities or nano-lithographic steps (Methods). The high density of NW cross-point junctions (~ $10^6$ NW junctions/mm$^2$) regulates the current/voltage distribution across the random network and ensures high connectivity of the system. While single crystalline Ag-NWs are highly conductive, the conductance of each NW junction is influenced by the mechanical stochasticity of the contact in between the crossed NWs and by the presence of an insulating polyvinylpyrrolidone (PVP) shell layer of ~ 1-2 nm surrounding the Ag-NW core (Extended Data Fig. 2). The PVP shell layer increases the NW junction resistance, thus representing one of the main issues for the realization of highly conductive electrodes based on Ag NWs[13,14]. However, PVP can be used as a solid electrolyte for exploiting cross-point junctions as electrochemical metallization cells[15]. We control the NWs network connectivity and its synaptic behavior employing two physical phenomena, different in nature:

i) By electrochemical potential difference applied between two intersecting NWs that induces anodic dissolution of Ag to form $Ag^+$ ions that migrate in the insulating shell layer under the action of the electric field and recrystallize to form a conductive bridge connecting the two NW cores (Fig. 1c). The formation/rupture of this conductive path is responsible for the observed memristive behavior of single NW cross-point junctions (Extended Data Fig. 3).

ii) Joule heating and electromigration driven electrical breakdown events occurring in single NWs at high current densities[16,17], responsible for the creation of a needle-like nanogap along the Ag-NWs (Fig 1d). Interestingly, the electrical connection can be regenerated by forming a conductive filament within the nanogap assisted by field-driven electromigration and bipolar electrode effects[18], as schematized in Fig. 1d. As a result, the nanogap induced by the breakdown of a single NW starts to behave as a memristive element (Extended Data Fig. 4).

In the above described framework, the NW random network represents a non-linear complex system that can be mapped onto a graph representation[19]. The NWs (nodes) are connected through memristive units (edges) as depicted in Fig. 1e. When a voltage difference is applied between any couple of nodes in the system, the current flowing in the network is distributed according to the Kirchhoff's current law and is regulated by the conductance (weight) of each memristive edge. The change of conductance in a single memristive edge is responsible for a redistribution of voltage/current across other nodes/edges of the network, inducing a cascade of conductance changes in other memristive edges through an avalanche effect that facilitates the emergence of spatially correlated structures of network activity. It is worth noticing that the structural topology can evolve depending on the network history, since breakdown events can divide single NW nodes into sub-nodes connected by a newly generated memristive edge.

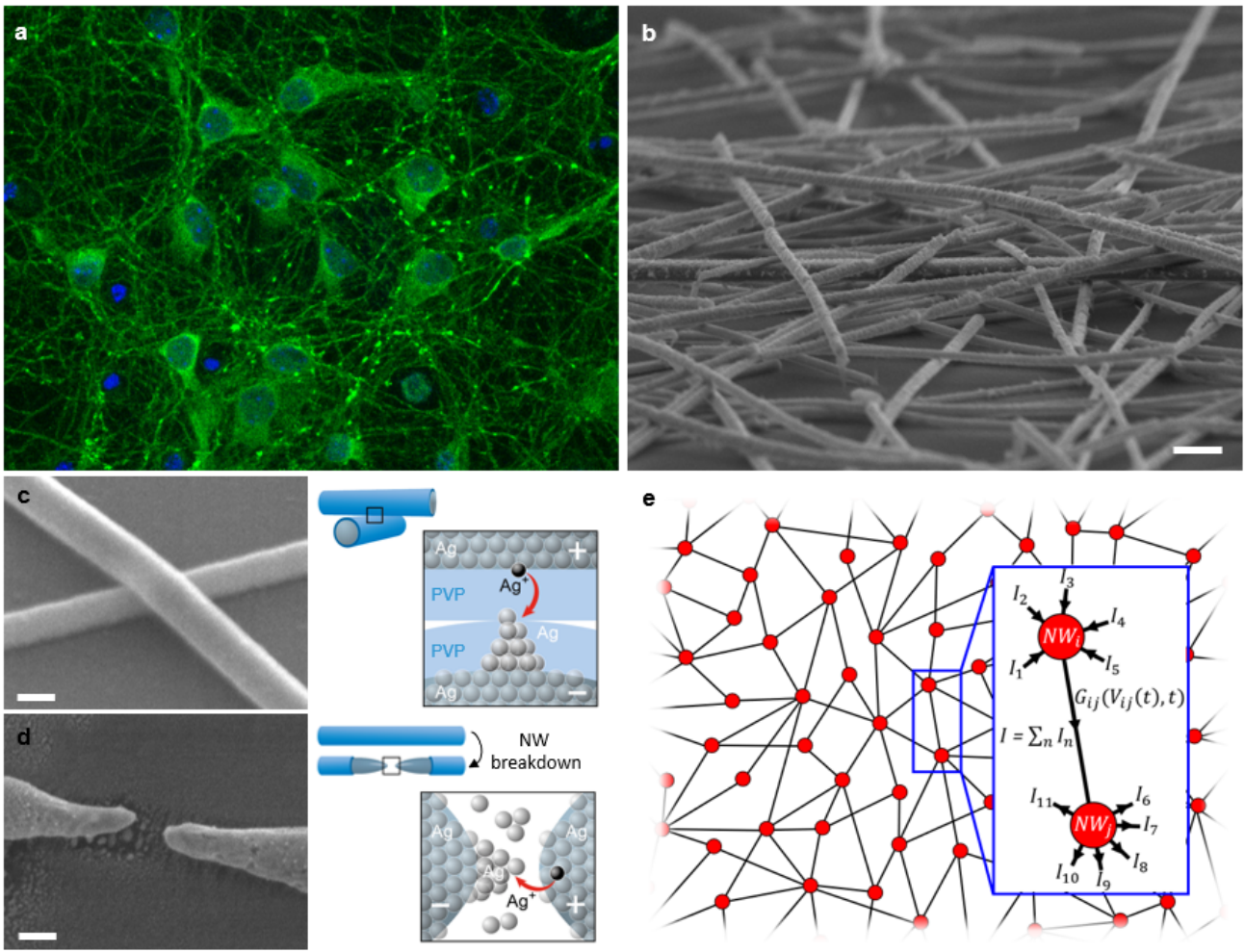

**Figure 1 | Neuromorphic hardware architecture of random Ag-NW networks. a** Biological neural network where the emergent behavior results from a multitude of synaptic interactions (bright fluorescent boutons) between neuronal processes and **b** a biologically inspired memristive NW network characterized by recurrent connectivity of Ag-NWs (scale bar, 500 nm). **c** SEM images and schematic representations of the memristive mechanism of a NW cross-point junction and **d** a current-induced nanogap formed after NW breakdown (scale bar, 100 nm). **e** Graph-theoretic abstraction of the NW network in which NW nodes (red dots) are connected through memristive edges (black lines). The enlarged view shows a detail of a memristive edge where the conductance $G$ depends on the voltage difference between the connecting nodes and on time, while the current flow is regulated by the Kirchhoff's current law.

The evolution of connectivity due to externally applied electric stimuli leads to a global memristive response of the network. After initialization (Extended Data Fig. 5a-b), the network measured in two-terminal configuration exhibits typical memristive behavior (pinched hysteresis loop) in the *I-V* plot (Fig. 2a). By applying a positive voltage sweep from 0 to 1 V (sweep 1), the network changed from an initial high-resistance state (HRS) to a low-resistance state (LRS) during the SET process. While the LRS is maintained during the voltage sweep from 1 to 0 V (sweep 2), voltage sweeps in the opposite polarity from 0 to -0.8 V (sweep 3) results in a RESET process turning the device to the initial HRS (sweep 4). The endurance characteristics, tested by switching the device 300 times between HRS and LRS by means of full-sweep cycles, reveals that the memristive behavior was maintained over cycling (Extended Data Fig. 5c). The time-evolution of the network stimulated by a constant sub-threshold bias voltage over large time scales revealed non-equilibrium dynamics with persistent conductance fluctuations and metastability. The power-law dependence of the intrinsic noise fluctuations in these non-equilibrium dynamic systems of highly interconnected nonlinear elements suggests distributed connectivity and self-organized criticality characterized by the spontaneous emergence of complexity from simple local interactions[20] (Extended data Fig. 6). Similar power-law scaling behavior was observed in spontaneous neural oscillations generated in the human brain[21]. More importantly, network dynamics can be exploited for the emulation of short-term synaptic plasticity (STP) that regulates the information exchange and processing within biological neural networks.[22,23] By applying an over-threshold constant voltage bias, the network conductance (synaptic weight) between two pads (neuron terminals) can be gradually increased (facilitation), as shown in Fig. 2b. The gradual enhancement of network connectivity is related to cascade switching events of memristive elements constituting the network that self-selects the lowest-energy path for electronic conduction[6]. After facilitation, the synaptic weight gradually relaxes back towards the initial state, exhibiting a volatile behavior due to the spontaneous dissolution of Ag conductive filaments previously formed in memristive elements of the network[24]. Note that an increase of the voltage bias stimulation results in a more intense potentiation of network conductivity, as well as in a longer relaxation time. The above-described network response to electrical stimuli allows the implementation of synaptic functionalities such as paired-pulse facilitation (PPF)[25,26]. Indeed, the network stimulation by means of short voltage pulses (mimicking action potentials) repetitively applied to the presynaptic pad with short inter-pulse time intervals results in a gradual increase of the network conductivity as a function of the number of applied pulses. The emergence of PPF during voltage pulse stimulation and the subsequent spontaneous relaxation of the network conductance are reported in Fig. 2c. The change in synaptic weight ($\Delta w$) and relaxation time can be modulated by changing the voltage pulse amplitude, with a higher voltage pulses resulting in larger changes of $\Delta w$ and longer relaxation times. Notably, PPF can be cyclically induced after device relaxation (Extended Data Fig. 7).

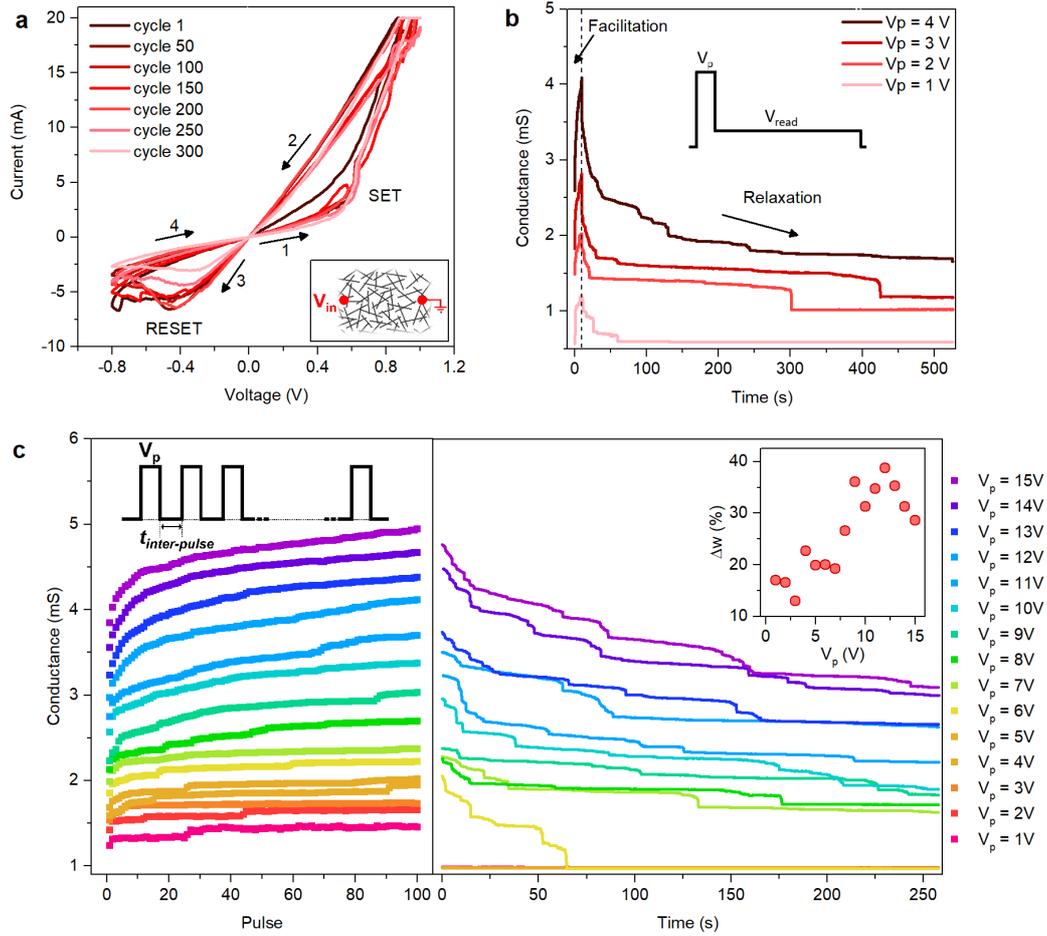

**Figure 2 | Memristive and neuromorphic functionalities of NW random networks. a** Resistive switching behavior of the NW network measured in a two-terminal configuration as schematized in the inset. **b** Gradual increase of conductance (synaptic facilitation) observed by applying a 10 s voltage pulse and subsequent conductance relaxation due to volatile resistive switching behavior read with a voltage of 50 mV. **c** Experimental demonstration of PPF in NW networks with gradual increase of conductance by stimulating the network with a train of 500 μs voltage pulses separated by 500 μs inter-pulse intervals (left panel) and relaxation process after stimulation (right panel) recorded by applying a constant voltage of 50 mV. The change of the synaptic weight (*Δw*) after 100 pulses is reported in the inset as a function of the applied pulse amplitude.

Due to the network connectivity, synaptic plasticity can be induced in each pair of nodes connected at least by one pathway of memristive edges. This means that a complex neural network composed of multiple synaptic connections can be reproduced by a multiterminal memristive network, in which each pad represents a "neuron terminal" (Fig. 3a and b). In this configuration, the synaptic weight between any given pair of "neuron terminals" is regulated by the global response of the network to external stimuli, endowing the system with intrinsic heterosynaptic plasticity. In biological systems, heterosynaptic plasticity is related to synaptic interactions that are responsible for a change in the strength of synapses that are not directly stimulated in addition to the specifically stimulated ones, providing distinct computational and learning properties to the network [12,27]. Previous attempts showed that such synaptic interactions can be emulated in top-down fabricated devices based on Ag nanoclusters[28] or 2D materials[29,30]. Here, Ag-NW network shows that the stimulation of a synaptic pathway (Extended Data Fig. 8) results in synaptic weight changes not only in the directly stimulated synapse but also in other nonstimulated synaptic pathways (Fig. 3c), thus mimicking heterosynaptic facilitation. In order to explain the intrinsic heterosynaptic behavior related to the functional connectivity of the NW-based system, we have developed a model that illustrates the network response to external electrical stimuli (Methods). The NW network is mapped into a grid graph of resistances and the experimentally observed potentiation of a directly stimulated synaptic pathway was modelled through the formation of a lower resistance path between the stimulated nodes (Fig. 3d). Modelling shows that the direct stimulation of a synapse results in a global redistribution of voltage/current across the network and brings about a change in the effective conductance of other synaptic pathways (Extended Data Fig. 9). In this framework, the synaptic network activity can be mapped onto a correlation map where each pixel represents the resistance (synaptic strength) between a specific couple of pads (neurons). Fig. 4a reports the correlation map of resistance variations ($\Delta R$) across the network after direct stimulation of the synapse connecting neuron terminals I and II. Besides a potentiation of the directly stimulated synaptic connection that exhibited the highest variation of resistance, the stimulation resulted also in changes of the strength of synaptic connections between nonstimulated neuron terminals. Note that experimental results are in qualitative accordance with model predictions (Fig. 4b). Interestingly, the change in synaptic strength in nonstimulated synapses depends on the spatial location of the corresponding terminals. Larger changes in the synaptic weight can be observed in synaptic pathways directly connected to previously stimulated terminals I or II (synapse I-III, I-IV, I-V, II-III, II-IV, II-V), while almost no changes are observed in other spatially distant synaptic pathways (synapse III-IV, III-V, IV-V). This is because the synaptic pathway resistance is not strongly influenced by a peripheral change in network connectivity. Instead, the stimulation of a more central synaptic pathway, such as synapse I-III (Fig. 4c and d), results in more generalized changes of synaptic weights in the system. These results, corroborated by considering also other pad configurations (Extended Data Fig. 10a-d), reveal a strong dependence of heterosynaptic changes on the spatial location of the primary plasticity effect. In addition, the short-term behavior of heterosynaptic facilitation was investigated by monitoring the evolution over time of synaptic weights after stimulation. Experimental data and modelling of the evolution of $\Delta R$ over time in synaptic pathways connecting terminal I to other terminals after direct stimulation of either synapse I-II or synapse I-III are reported in Fig. 4e-f and 4 g-h, respectively. Time course of synaptic weights after different stimuli are reported in Extended Data Fig. 10e-h. Results demonstrated that the device exhibits short-term heterosynaptic plasticity, with the strength of all synaptic pathways tends to restore to the initial conditions over time due to the previously discussed memristive network relaxation. Apart from experimental fluctuations, experiment and model clearly show a similar spatio-temporal correlation.

In conclusion, the electrochemically controlled connectivity of the memristive random NW network exhibits functional and structural plasticity that mimics the behavior of biological neural circuits by displaying homo- and hetero-synaptic plasticity and activity-dependent changes in the connectivity map. In contrast to conventional neural networks realized with a top-down approach based on memristive devices or transistors, these biologically inspired systems allow a low-cost realization of neural networks fabricated through a bottom-up approach that can learn and adapt when subjected to external stimuli, strictly mimicking the processes of experience-dependent synaptic plasticity that shape the connectivity and functionalities of the nervous system. These results represent a radically new approach toward the development of biologically inspired intelligent systems able to physically compute data from multiple inputs through the Kirchhoff's laws. Furthermore, they highlight the way for the hardware implementation of unconventional computing paradigms, such as reservoir computing, where an input signal has to be mapped into a higher dimensional output.

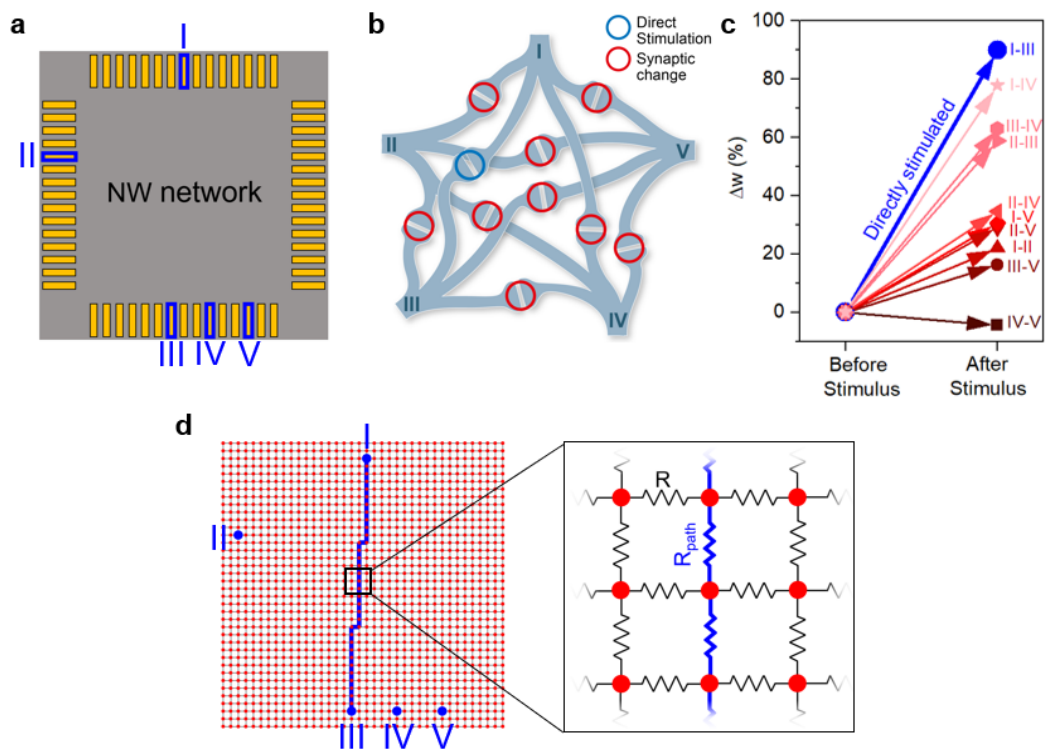

**Figure 3 | Implementation of heterosynaptic plasticity in multiterminal memristive networks. a** Schematization of the multiterminal memristive NW network device with highlighted pads used during heterosynaptic experimental demonstration and **b** corresponding biological representation of the device, where synaptic interactions lead to heterosynaptic plasticity characterized by a change of the synaptic weight of synapses that are not directly stimulated. **c** Direct stimulation of synapse I-III in the NW network results in synaptic weight changes not only in the directly stimulated synapse but also in other nonstimulated conductive pathways. **d** Modelling of the system through a grid graph of resistances, with the blue path representing the stimulated synaptic pathway.

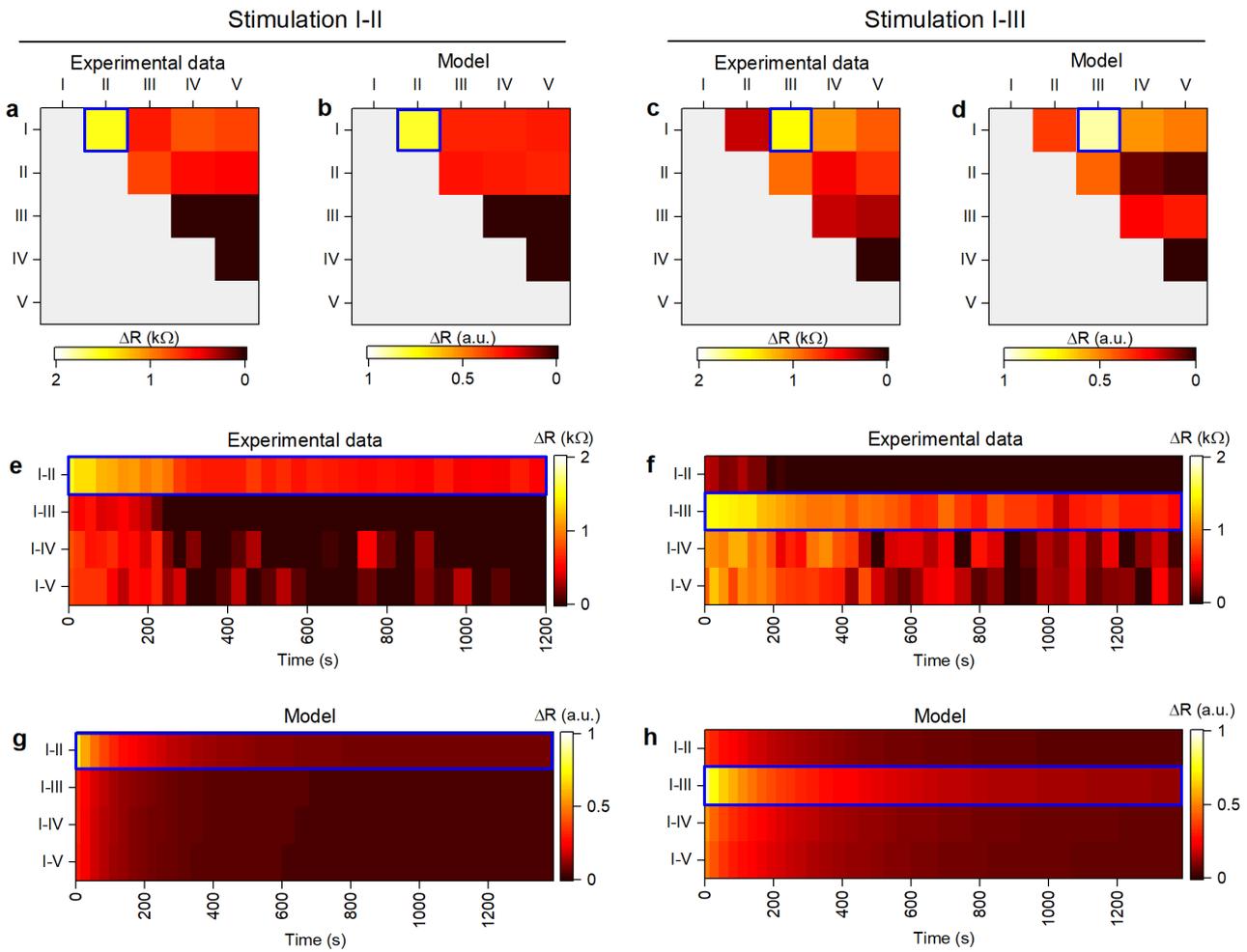

**Figure 4 | Experimental demonstration and modelling of heterosynaptic plasticity.** Experimental and simulated correlation maps of resistance variation (*ΔR*) of synaptic pathways after stimulation of synapse I-II (**a** and **b**) and stimulation of synapse I-III (**c** and **d**). Experimental data of the relaxation process over time of synaptic pathways connecting synapse I after stimulation of synapse I-II (**e**) and synapse I-III (**f**) with corresponding simulated data in panel **g** and **h**, respectively. In all panels, directly stimulated synaptic pathways are highlighted in blue.

**Methods**

**NW characterization**

Ag-NWs with diameter of 115 nm and length of 20-50 μm in isopropyl alcohol suspension were purchased from Sigma-Aldrich. Chemical and structural characterization of NWs (Extended Data Fig. 2) evidenced the presence of a polyvinylpyrrolidone (PVP) layer surrounding the Ag NW core. The presence of this shell layer is an unavoidable consequence of the polyol synthesis process, since this polymer is commonly used as surfactant to control the growth rates of selected planes of face-center-cubic silver in order to obtain the NW morphology[31,32]. Structural characterization of NWs was performed by Transmission Electron Microscopy (TEM) using a FEI Tecnai F20ST equipped with a field emission gun (FEG) operating at 200 kV. The sample was prepared by drop casting NWs in suspension onto a TEM grid. Chemical composition of Ag-NWs dispersed on a $SiO_2$ substrate was assessed by X-ray photoelectron spectroscopy (XPS), using a Kα source with energy of 1486.6 eV and using the C 1s peak position (284.8 eV) as calibration. All subpeaks of C 1s and O 1s were fitted by means of Gaussian−Lorentzian functions after Shirley's background subtraction.

**Fabrication of NW memristive networks**

NW random networks were fabricated by drop casting and spontaneous solvent evaporation of Ag-NWs in isopropyl alcohol suspension (~ 0.13 %) on a $SiO_2$ (1 μm)/Si commercial substrate. The network structural topology was characterized using field emission scanning electron microscopy (FE-SEM; Zeiss Merlin). To investigate the response of the network to electrical stimuli, metallic Au pads (thickness of 150 nm) were realized on the NW network by sputtering and shadow mask. Multi-terminal planar devices schematized in Fig. 3a are made of 60 pads (approximate size of 1.3 x 0.3 mm) organized around a 7 x 7 mm square. The distance between adjacent pads is 100 μm.

**Fabrication of single cross-point NW junctions and single NW devices**

To fabricate devices based on single cross-point NW junctions and on single NWs (Extended Data Fig. 3a and 4a, respectively), Ag-NWs were distributed by drop casting on an insulating $SiO_2$ substrate pre-patterned with a sub-millimeter probe circuit realized by direct laser writing lithography and Ti/Au deposition. Then, selected single cross-point NW junctions or single NWs were connected to the probe circuit by Pt deposition through an Ion Beam Induced Deposition (IBID) with a gas injection system (GIS) in a FEI Quanta™ 3D Microscope. Note that the realization of contacts by this technique i) ensures direct contact between the deposited Pt and the Ag-NW core since the few-nanometer thick PVP coating shell-layer is easily removed in the Pt/Ag contact area during ion beam-induced deposition of Pt and ii) avoid any interaction of NWs with solvents and polymers necessarily employed during conventional Electron Beam Lithography (EBL) that can interact with the PVP coating layer and alter the original structure of the NW.

**Two-terminal electrical measurements**

Electrical characterizations of single cross-point NW junctions, single NW devices and NW networks in two-terminal configuration were performed by using a Keithley 4200 semiconductor device analyser equipped with Pulse Measuring Units (PMUs) and coupled with a SemiProbe probe station. Electrical characterization of NW networks in two-terminal configuration (Fig. 2) were performed by considering Au electrodes separated by 7 mm. *I-V* cycles reported in Fig. 2a were performed by applying a voltage sweep of 0.27 V/s and by externally imposing a compliance current (CC) of 20 mA. The analysis of power spectral density of noise under constant bias stimulation reported in Extended Data Fig. 6 was performed by means of Fast Fourier transform with normalized power to Mean Square Amplitude. The change in the synaptic weight (*Δw*) reported in Fig. 2c was calculated as *Δw=[G(n)- G(1)]/ G(1)*, where *G(n)* is the conductance of the synaptic pathway during the last pulse of the applied pulse train *(n=100)*. All measurements were performed in air at room temperature.

**Multi-terminal characterization**

For multi-terminal characterization, the device was arranged in a conventional probe station equipped with multiple electrical probe tips controlled by micromanipulators. A Keithley 707 switch matrix was connected to the probes and the instruments to correct routing the measurements, allowing a sequential selection of each combination of pad pairs, while keeping other pads floating. The electrical characterization was conducted by using a TTI-TGA 1202 arbitrary waveform generator with a 100 Mhz bandwidth able to deliver rectangular voltage pulses of various duration and amplitude, while the current flowing into the sample was collected by a Lecroy Wavesurfer 3024 oscilloscope with 200MHz bandwidth and 4 GSample/s sampling rate. All measurements were controlled with GPIB connection by a remote desktop PC. Heteroplasticity was investigated by monitoring the resistance evolution in between all the considered synaptic pathways after direct stimulation of a selected synapse. Direct stimulation of a selected synapse was performed by applying a 1-s voltage pulse of 8 V at its terminals (Experimental Data Fig. 8) while electrical resistance between each couple of pads was read by applying a 100 μs voltage pulse with amplitude of 0.1 V. Correlation maps of resistance variation ($\Delta R$) reported in Fig. 4 a-d and Extended data Fig. 10 a-d were obtained by comparing resistance maps before and after the direct stimulation of a selected synapse. The resistance variation was calculated as $\Delta R = -(R_{post\_stimulus} - R_{pre\_stimulus})$ and the variation of $\Delta R$ has to be intended as a decrease of resistance of the synaptic pathway after stimulation. The heterosynaptic relaxation process reported in Fig. 4 e-h and Extended data Fig. 10 e-h was recorded by monitoring the evolution of $\Delta R$ over time of the selected synaptic pathways. Note that color maps limits were restricted to positive $\Delta R$ values for better data visualization, experimental data with $\Delta R < 0$ arising from experimental data fluctuations were considered as tail values.

**Modelling the multi-terminal NW network device**

Simulation of the network response was performed in Python employing the NetworkX package by modelling the memristive NW network with a square grid of resistances (Fig. 3 c). The resistive network, composed of 38 x 38 nodes, was generated by connecting each pair of adjacent nodes with a resistive edge $R$. The distance between each node pair corresponds to about 200 μm of the NW random network. Nodes in particular spatial locations were selected as input/output nodes (pads) to mimic the geometry of the NW network device. The effective resistance in between two pad nodes of the grid was calculated using nodal voltage analysis and solving nodal equations of Kirchhoff's first law to find voltage at each node of the circuit. To perform these calculations, a read current was imposed to flow in one pad node, while keeping the other pad node at ground (Extended Data Fig. 8b and c). Direct stimulation of a synaptic pathway was modelled through the formation of a low resistance conductive path (shortest path) connecting nodes in between its terminals, as reported in Extended Data Fig. 9d. For this purpose, the resistance value of edges of this path was set to $R_{path}$. This assumption is in accordance with previous simulations that reported the formation of a "winner-takes-all" conductive pathway in Ag-NW networks[6,33]. Finally, the current intensity and direction in a generic edge connecting nodes $i$ and $j$ was calculated as $V_{ij}/R_{ij}$, where $V_{ij}$ is the voltage difference between the connecting nodes and $R_{ij}$ is the edge resistance. During simulations, the effective resistance between two pad nodes of the network was calculated by injecting in the input pad node a current of 0.001 a.u., considering $R$=1000 a.u. and $R_{path}$= R/10. The heterosynaptic relaxation process over time was simulated by progressively increasing the resistance of edges composing the conductive path ($R_{path}$). During the relaxation process, $R_{path}$ was increased according to an exponential function $R_{path}(t)=A-(A-R_{path}(t=0))\cdot exp(-t/\tau)$, where $R_{path}(t=0)=R/10$, while parameters $A$ and the decay time constant $\tau$ were obtained from interpolation of experimental data of the decay process of the directly stimulated synapse. In Extended Data of Fig. 10 e and g, $R_{path}$ was decreased linearly over time since experimental data were well interpolated by means of linear regression. In this fashion, it was possible to simulate the progressive dissolution of the conductive path formed during stimulation and follow over time the variation of other nonstimulated pathways

**Data availability**

The data that support the findings of this study are available from the authors on reasonable request, see author contributions for specific data sets.

**Acknowledgements**

The support by Mauro Raimondo in helping with SEM measurements, by Katarzyna Bejtka for performing TEM measurements, by Salvatore Guastella for performing XPS measurements, and by Thomas Poessinger for helping with graphics is gratefully acknowledged. Device fabrication was performed at "Nanofacility Piemonte", a facility supported by the "Compagnia di San Paolo" foundation.

**Author Contributions**

G.M., I.V. and C.R. generated the idea and designed the experiments. G.M and M.F. performed device fabrication and characterization. G.P. and G.M. performed multielectrode electrical measurements. G.M., G.P. and D.I. developed the electrical model. G.M., G.P., D.I. and C.R. analysed the data; F.B. provided the neurological interpretation of the results. L.B., D.I., I.V. and C.R. supervised the research. All authors participated in the discussion of results and revision of the manuscript.

**Competing interests**

The authors declare no competing financial or non-financial interests.

**Self-organizing memristive nanowire networks with structural plasticity emulate biological neuronal circuits - Extended Data**

Gianluca Milano[1,2], Giacomo Pedretti[3], Matteo Fretto[2], Luca Boarino[2], Fabio Benfenati[4,5], Daniele Ielmini*[3], Ilia Valov*[6,7], Carlo Ricciardi*[1]

[1]Department of Applied Science and Technology, Politecnico di Torino, C.so Duca degli Abruzzi 24, 10129 Torino, Italy.

[2]Advanced Materials Metrology and Life Science Division, INRiM (Istituto Nazionale di Ricerca Metrologica), Strada delle Cacce 91, 10135 Torino, Italy.

[3]Dipartimento di Elettronica, Informazione e Bioingegneria, Politecnico di Milano and IU.NET, Piazza L. da Vinci 32, 20133, Milano, Italy.

[4]Center for Synaptic Neuroscience and Technology, Istituto Italiano di Tecnologia, Largo Rosanna Benzi, 10, 16132 Genova, Italy.

[5]IRCCS Ospedale Policlinico San Martino, Largo Rosanna Benzi, 10, 16132 Genova, Italy.

[6]JARA – Fundamentals for Future Information Technology, 52425 Jülich, Germany.

[7]Peter-Grünberg-Institut (PGI 7), Forschungszentrum Jülich, Wilhelm-Johnen-Straße, 52425 Jülich, Germany.

*e-mails: daniele.ielmini@polimi.it; i.valov@fz-juelich.de; carlo.ricciardi@polito.it;

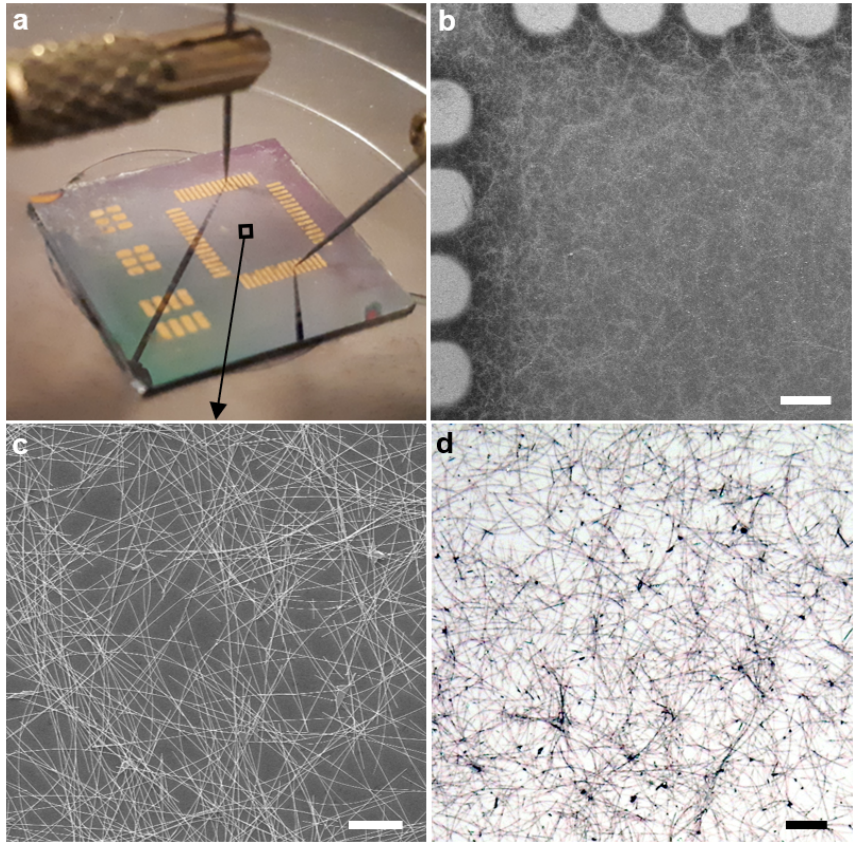

**Extended Data Figure 1 | Multi-terminal memristive device based on NW networks. a** Multi-terminal memristive device during electrical measurements in two-terminal configuration where electrical probes contact the Au electrodes. **b** SEM image of the NW-based device showing Au electrodes deposited on the NW network (scale bar, 250 μm). **c** SEM image (scale bar, 10 μm) and **d** optical image (scale bar, 50 μm) showing an enlarged view of the highly interconnected NW network topology.

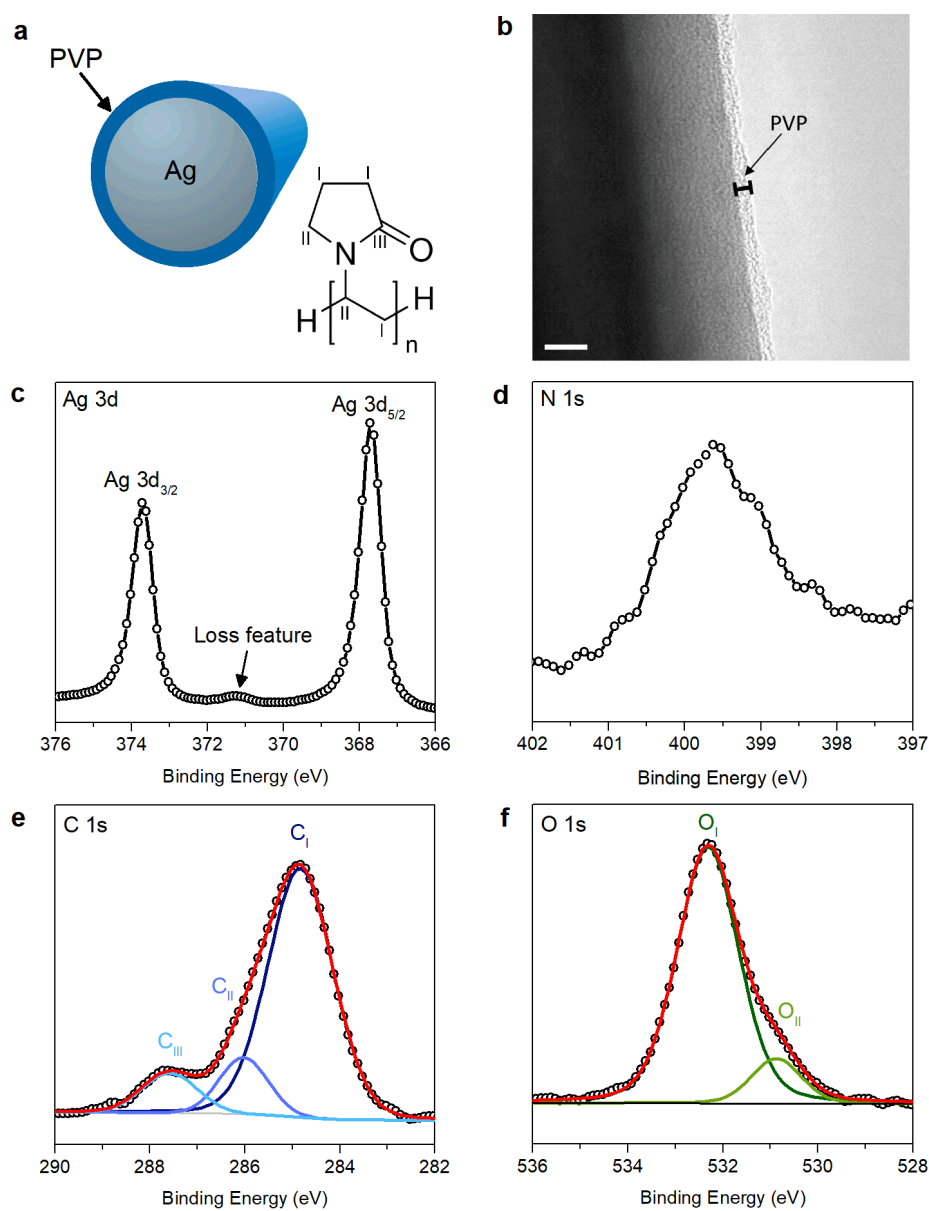

**Extended Data Figure 2 | Structural and chemical characterization of PVP coated Ag nanowires. a** Schematic representation of an Ag-NW surrounded by a Polyvinylpyrrolidone (PVP) coating layer. The PVP molecule contains carbon species with different chemical environments that can be labelled as I, II and III in the scheme. **b** TEM characterization of a crystalline Ag-NW revealing the presence of the amorphous PVP layer with thickness of 1-2 nm (scale bar, 5 nm). The presence of a PVP coating layer on the Ag nanostructures was further confirmed by means of XPS measurements, where binding energy was referenced to the standard C 1s at 284.8 eV. **c.** High resolution XPS spectra of the Ag 3d core level, revealing the presence of doublet spectral lines located at 373.7 eV and 367.7 eV attributable to Ag $3d_{3/2}$ and Ag $3d_{5/2}$ electrons, respectively, with a

spin-orbit splitting of 6 eV. **d** The presence of the N 1s peak located at about 399.6 eV represents a fingerprint of the presence of a PVP coating layer. **e** In accordance to the PVP molecular structure shown in panel a, the deconvolution of the C peak revealed the presence of three carbon species with different chemical environments: the prominent $C_I$ component located at 284.8 eV can be attributed to the C-C bonds, while the $C_{II}$ and $C_{III}$ components located at 286.0 eV and 287.6 eV can be attributed to the C-N and C=O bonds, respectively. **f** The O 1s peak can be deconvoluted in two components $O_I$ and $O_{II}$ located at 532.3 eV and 530.9 eV, respectively. The $O_I$ component is associated with oxygen in the carboxyl group of PVP while the $O_{II}$ component can be attributed to oxygen species in the carboxyl groups that are interacting with the Ag NW surface.

.

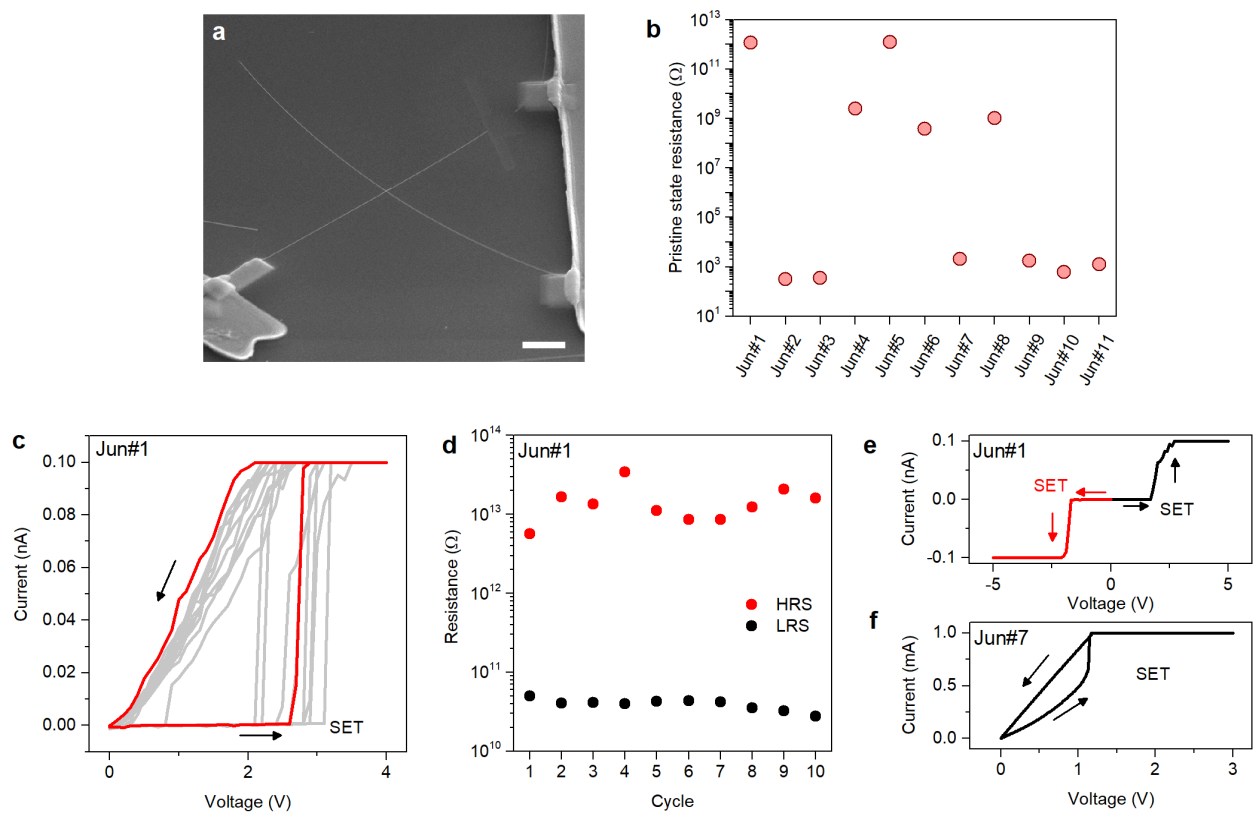

**Extended Data Figure 3 | Memristive behavior of single NW junctions. a** SEM image of a single NW cross-point junction device realized by IBID technique (scale bar, 5 μm). **b** A wide range of pristine state resistances were observed by considering various NW cross-point junctions due to the stochasticity of the mechanical contact between NWs at the intersection with the presence of the PVP coating layer. While part of the junctions is in a low resistance state suggesting a good mechanical contact between the Ag-NW inner cores of the intersecting nanowires, others are in a high resistance state. **c** Resistive switching behavior of a single NW cross-point junction that is initially in a high resistance state. Note that the resistive switching behavior is volatile, since the device spontaneously relaxed back to the high resistive state after a SET event without the need of a RESET process (consecutive voltage sweeps were performed with a delay time of 60 s). **d** Repeated cycling of the single NW cross-point junction (read voltage of 0.6 V). **e** The SET event can be induced in both voltage polarities due to the symmetric structure of the cross-point memristive cell. **f** Resistive switching behavior of a single NW cross point junction characterized by a lower pristine

state resistance. Despite the larger current involved, also in this case device stimulation with a voltage sweep resulted in a resistance decrease.

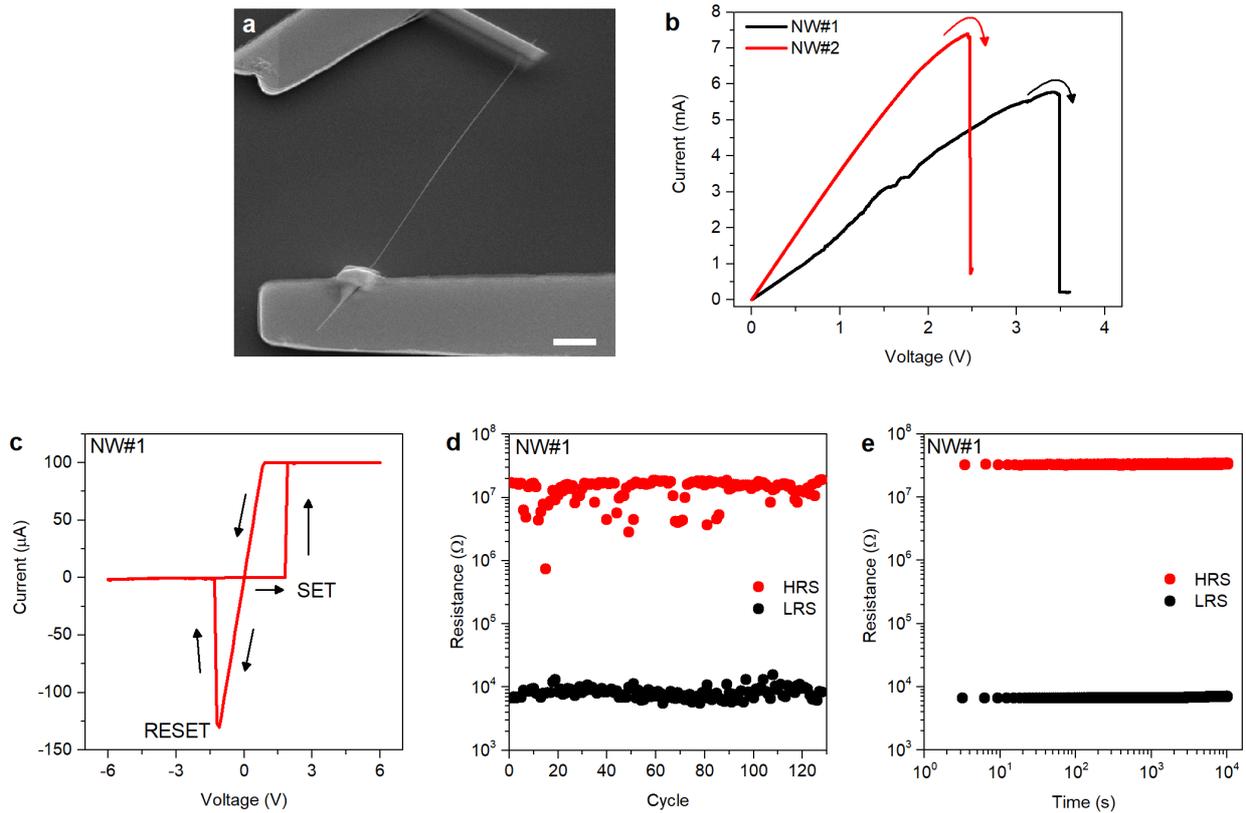

**Extended Data Figure 4 | Memristive behavior of single NWs. a** SEM image of a single NW device realized by IBID technique (scale bar, 5 μm). **b** Electrical breakdown of single Ag-NWs. By applying a voltage sweep to the nanostructure that initially exhibited high conductivity, a sudden drop of current due to NW electrical failure can be observed. The breakdown of single Ag-NWs was observed to occur in correspondence of an injected electrical power of ≈ 20 mW. **c** The breakdown-induced nanogap (see Fig. 1 d) acts as a resistive switching element characterized by non-volatile bipolar resistive switching behavior. **d** Endurance of the memristive cell acquired by full-sweep cycles (read voltage of 0.2 V). **e**, Retention of the device after being switched to the low resistance state (LRS) and high resistance state (HRS), acquired by applying a constant stress voltage of 50 mV.

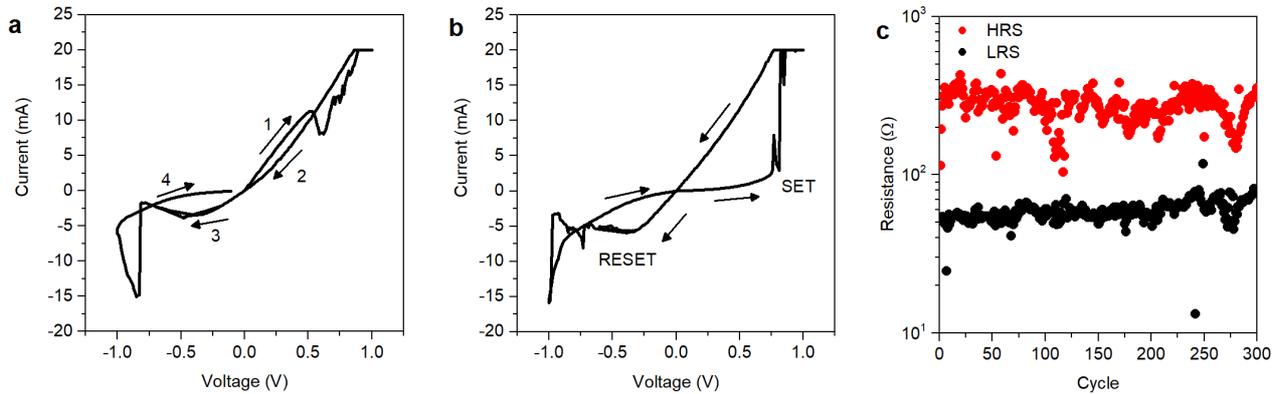

**Extended Data Figure 5 | Initialization and endurance properties of the memristive network.**
**a** Pristine state resistance of the network depends on the structural topology, as well as on the pristine state resistance of junctions along the numerous percolative paths that are connecting the involved electrode pads. Before exhibiting stable resistive switching behavior, a voltage sweep applied to the network resulted in numerous SET/RESET transitions attributable to an initial assessment of memristive elements composing the network. Typically, the first voltage sweep (sweep 1) resulted in a RESET process that can be attributed to the annihilation of an initial conductive percolative path with consequent redistribution of the current flow across the network. This behavior can be attributed to current-induced NW breakdown events. **b** After initialization, the network stimulated with voltage sweeps exhibited bipolar resistive switching behavior. **c** The endurance properties of the device acquired by full-sweep cycles (see Fig. 2a) reveal that the device maintained a memristive behavior over cycling.

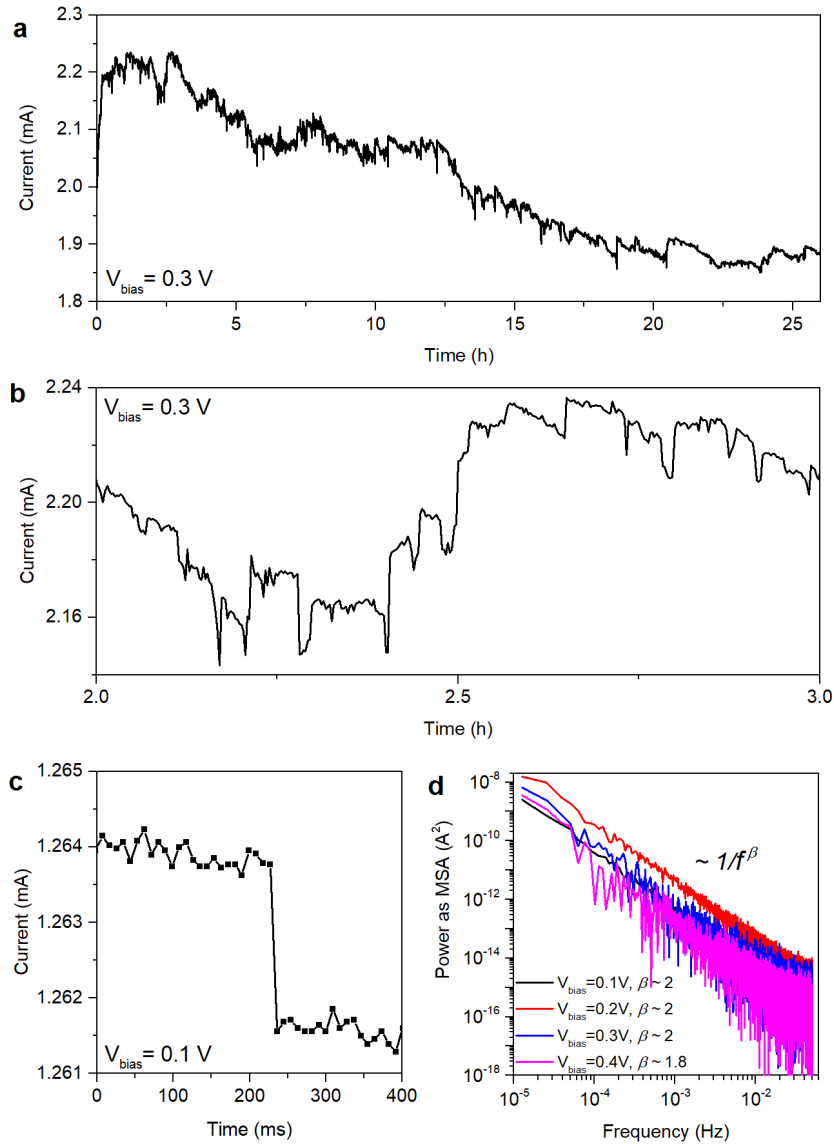

**Extended Data Figure 6 | Intrinsic noise in random NW networks. a** Persistent fluctuation and temporal metastability of network conductivity over large time scales under constant subthreshold voltage bias stimulation (0.3 V). **b** Enlarged view of part of the current time trace presented in panel a, showing a detail of the step-like current fluctuations ascribed to the continuous reorganization of network connectivity. **c** Detail of a single step-like current transition of the network in the millisecond timescale observed under constant bias stimulation of 0.1 V. **d** The analysis of the power spectral density of network noise, obtained by performing the Fourier Transform of time

traces of current responses at various constant voltage biases, evidenced a $1/f^\beta$ power-law trend of fluctuations with $\beta \approx 1.8\text{-}2$. Current traces over time for each considered voltage bias were recorded for 22 h.

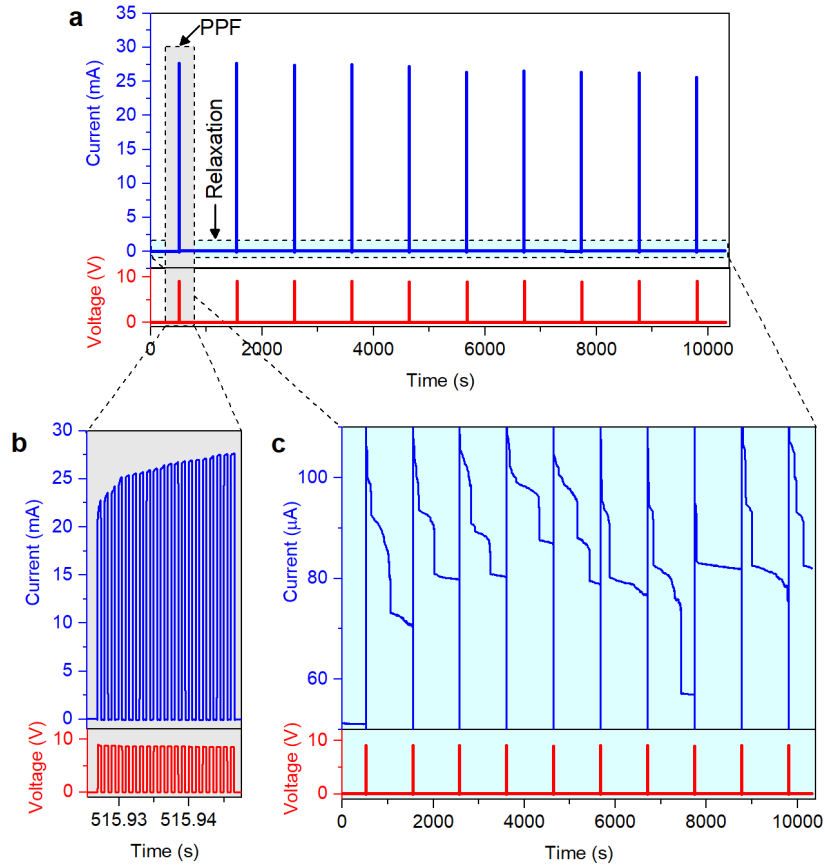

**Extended Data Figure 7 | Paired-Pulse Facilitation (PPF) and relaxation cycles. a** Repeated PPF and subsequent relaxation of the stimulated synaptic pathway. The device was cyclically stimulated with trains of voltage pulses (10 V, 500 μs pulses, 500 μs inter-pulse intervals, 20 pulses), while the relaxation process after stimulation was recorded at a reading voltage of 50 mV. **b** Enlarged view of the current response of the device to voltage pulses during PPF. The application of voltage pulses resulted in a progressive increase of current due to the gradual increase of network conductance. **c** Enlarged view of the step-like relaxation processes of the network after PPF events.

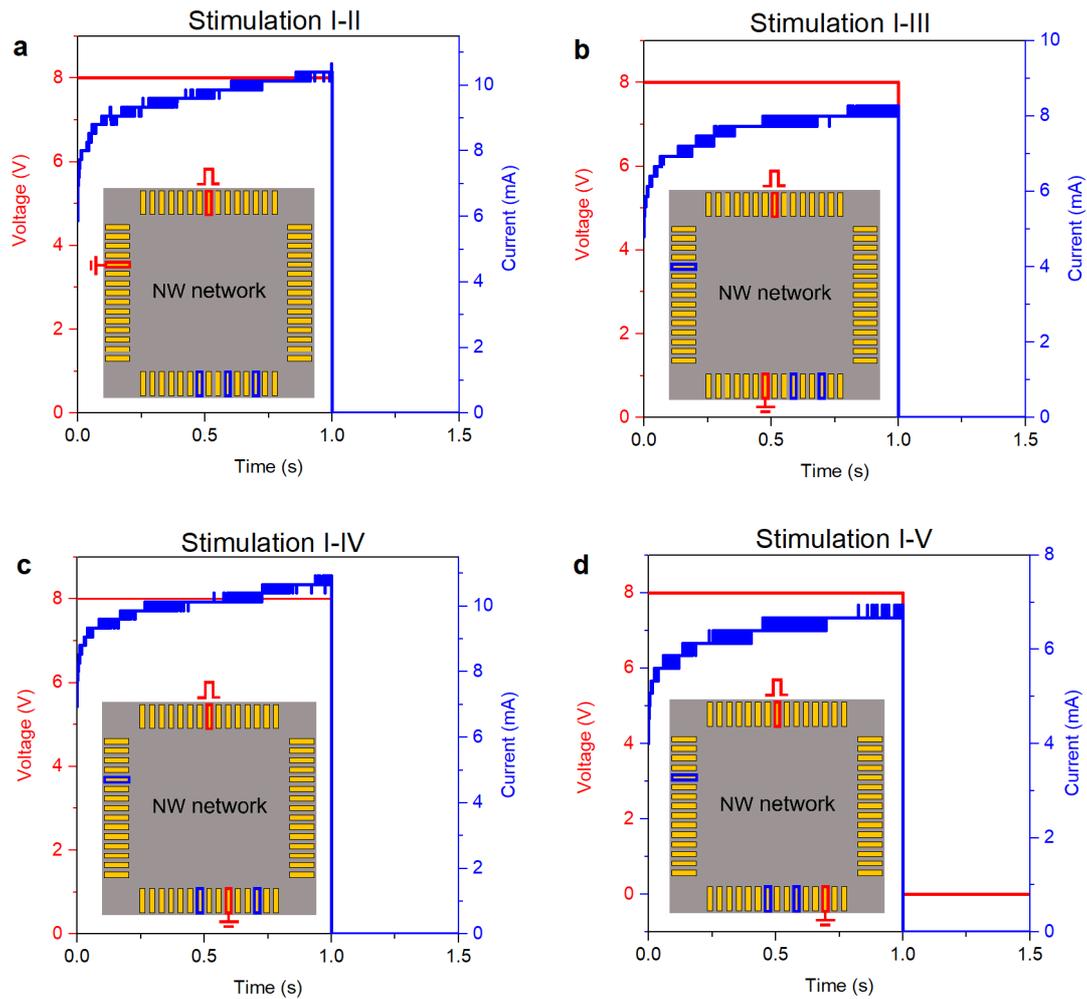

**Extended Data Figure 8 | Experimental stimulation of different neuron terminal pairs.** Current response of the synaptic pathway during stimulation of pad terminals **a** I-II, **b** I-III, **c** I-IV and **d** I-V. Stimulation was performed by applying an 8 V pulse of 1 s between electrode I and the other selected electrode (II, III, IV, V) that was fixed at ground, keeping floating all the other electrodes. A schematic representation of the electrical connections during synaptic stimulation is reported in the insets, where stimulated pads are highlighted in red, while other considered electrodes are in blue. In all cases, a gradual increase of current was observed because of the progressive reinforcement of the directly stimulated synapse.

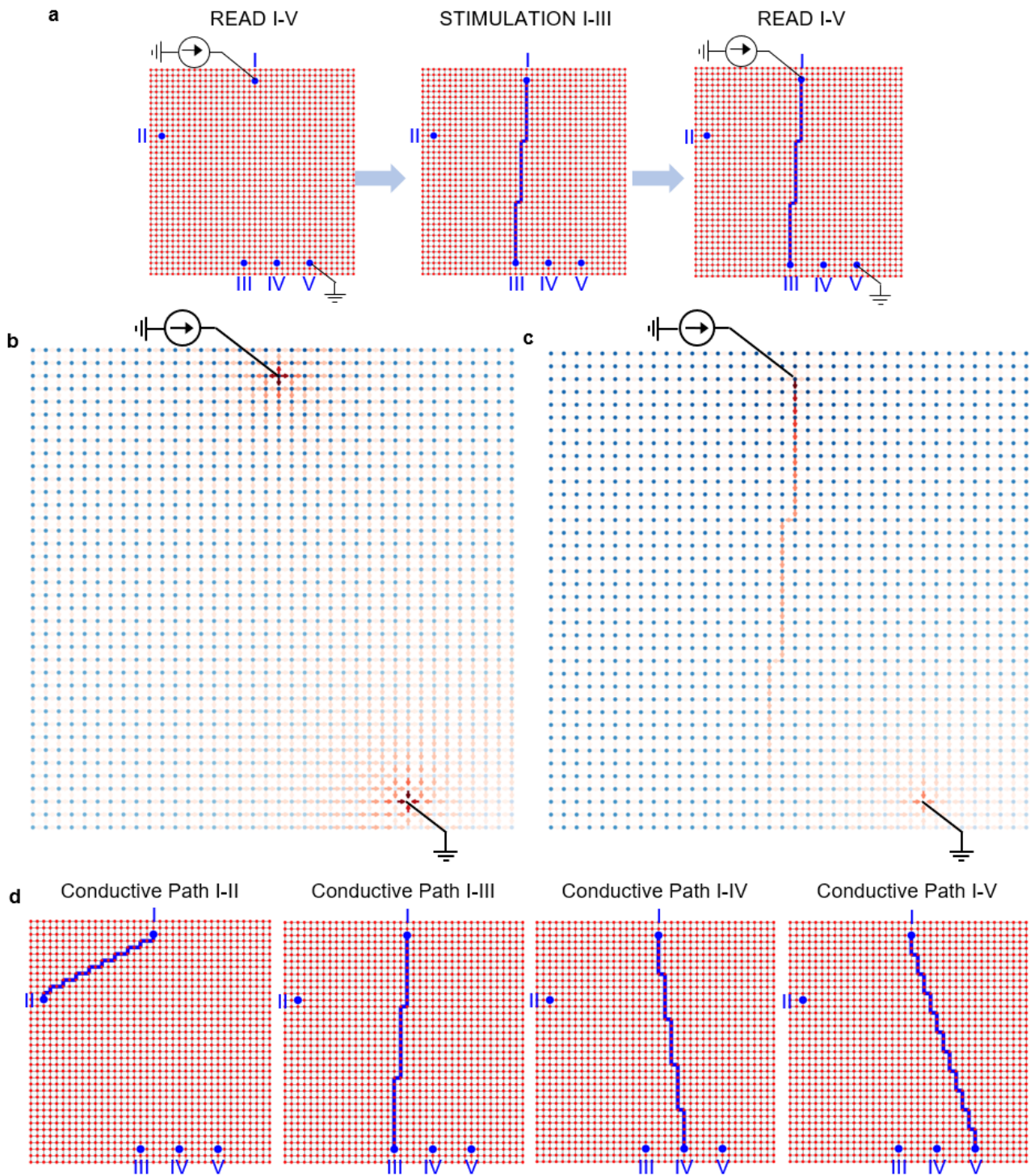

**Extended Data Figure 9 | Current/voltage redistribution across the network after synapse stimulation. a** Schematic flow used for modelling the redistribution of voltage/current across the network after stimulation of a synapse. Changes in the effective resistance of nonstimulated synapses were probed by comparing their resistance before and after the formation of a low-

resistance conductive path (blue path) between stimulated neuron terminals. Current/voltage distribution across the network during the read process of synapse I-V before (**a**) and after (**b**) direct stimulation of synapse I-III. The red color intensity of arrows, indicating the direction of the current flow, is proportional to current intensity, while the blue color intensity of nodes is proportional to the node voltage. During simulation, current was injected into pad I, while pad V was fixed at ground. **d** Complete set of conductive pathways (shortest paths) exploited for the simulation of the synapse stimulation.

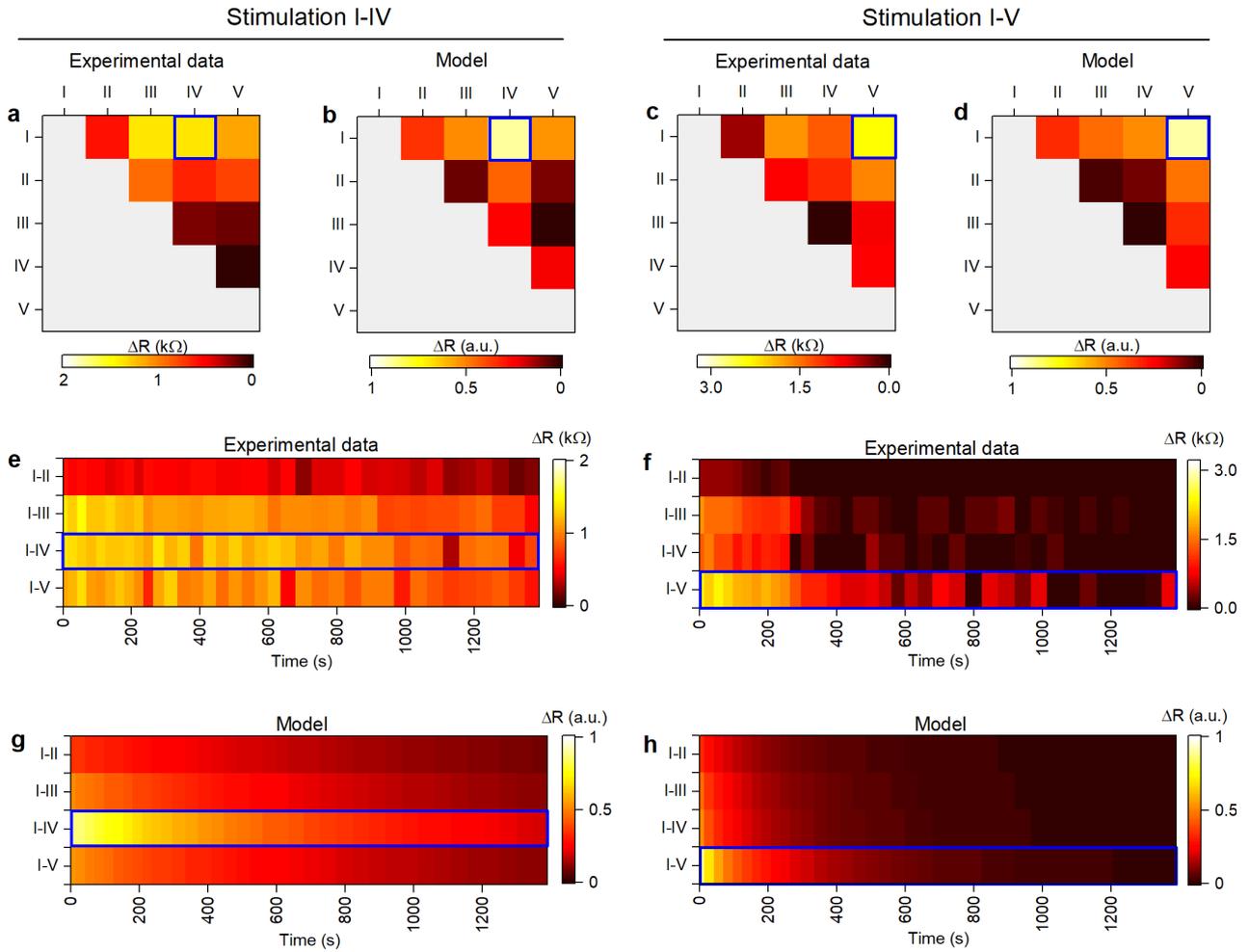

**Extended Data Figure 10 | Heterosynaptic plasticity in different configurations.** Experimental and simulated correlation maps of resistance variation ($\Delta R$) of synaptic pathways after stimulation of synapse I-IV (**a** and **b**) and of synapse I-V (**c** and **d**). Experimental data of the relaxation process over time of synaptic pathways connecting synapse I after stimulation of synapse I-IV (**e**) and synapse I-V (**f**) with corresponding simulated data in panel **g** and **h**, respectively. In all panels, directly stimulated synaptic pathways are highlighted in blue.